\documentclass[nofootinbib,amsmath,amssymb]{revtex4}

\usepackage{amsfonts}
\usepackage{bm}
\usepackage{color}
\usepackage{graphicx}
\usepackage{amsmath}
\usepackage{amsfonts}
\usepackage{amssymb}
\usepackage{color}
\usepackage{epsf}

\newcommand{\be}{\begin{equation}}
\newcommand{\ee}{\end{equation}}
\newcommand{\ba}{\begin{eqnarray}}
\newcommand{\ea}{\end{eqnarray}}
\newcommand{\nn}{\nonumber}

\newcommand{\no}{\nonumber}

\def\VEV#1{\left\langle #1 \right\rangle}

\newcommand{\dispersion}{\VEV{\left(\Delta \vec v\right)^2}}

\begin{document}

\title{Constraining Dark Matter-Baryon Scattering with Linear Cosmology}

\author{Cora Dvorkin}
\altaffiliation{cdvorkin@ias.edu}
\author{Kfir Blum}
\altaffiliation{kblum@ias.edu}
\affiliation{Institute for Advanced Study, School of Natural Sciences, Einstein Drive, Princeton, NJ 08540, USA}
\author{Marc Kamionkowski}
\altaffiliation{kamion@jhu.edu}
\affiliation{Department of Physics and Astronomy, Johns Hopkins University, Baltimore, MD 21218, USA}

\begin{abstract} 
We derive constraints on elastic scattering between baryons and
dark matter using the cosmic microwave background (CMB) data from the {\it Planck satellite} and the
Lyman-$\alpha$ forest data from the {\it Sloan Digital Sky Survey}. Elastic scattering allows baryons and
dark matter to exchange momentum, affecting the
dynamics of linear density perturbations in the early
Universe.  
We derive constraints to scattering cross sections of the form $\sigma
\propto v^n$, allowing for a wide range of velocity dependencies with $-4\leq n \leq 2$. We improve and correct previous
estimates where they exist, including velocity-independent cross section as well as dark matter millicharge and electromagnetic dipole moments. Lyman-$\alpha$ forest data dominates the
constraints for $n>-3$, where the improvement over CMB data alone can be several orders of magnitude. Dark matter-baryon scattering cannot affect the halo mass function on mass
scales $M>10^{12}\,M_\odot$. Our results imply, model-independently, that a baryon in the halo of a galaxy like our own Milky Way, does not scatter from dark matter particles during the age of the galaxy. 
\end{abstract}

\maketitle

\tableofcontents
\section{Introduction}
\label{sec:intro}

The canonical dark matter (DM) candidate is assumed to interact with
Standard Model particles only gravitationally.  The cross
section for the simplest weakly-interacting massive particles
(WIMPs)
\cite{Jungman:1995df,Bergstrom:2000pn,Bertone:2004pz,D'Amico:2009df}
to scatter from baryons is non-zero but sufficiently
small to be considered effectively zero for scales above a solar mass~\cite{Loeb:2005pm}. However, the shrinking
of the canonical-WIMP parameter space from null LHC and direct
searches, efforts to explain the coincidence between
the DM and baryon densities \cite{Kaplan:2009ag,Zurek:2013wia},
as well as possible difficulties for collisionless N-body simulations to reproduce observational data~\cite{Moore:1997jy,Kamionkowski:1999vp,Bullock:2010uy,Weinberg:2013aya,Famaey:2013ty,Peebles:2013hla}, provide motivation to consider stronger baryon-DM interactions.

In this paper we evaluate the constraints to elastic DM-baryon
scattering that arise from the cosmic microwave background
(CMB) and large-scale structure (LSS).  In the standard scenario
of collisionless DM, perturbations to the DM density grow in
amplitude early on while the pressure in the
baryon-photon fluid prevents it from falling into the
DM-dominated potential wells.  This dynamics defines the overall shape of the CMB and matter power spectra, and gives rise to characteristic features such as the
acoustic peaks seen in the CMB power spectrum and the baryon
acoustic oscillations (BAO) in the matter power spectrum.  If
there is some coupling between DM and baryons, then the
drag force between the baryon-photon fluid and the DM affects
the baryon-photon oscillations and suppresses the growth of
perturbations to the dark matter, and hence total, density.  The
beautiful agreement between the predictions of the colisionless $\Lambda$CDM
model and the wealth of CMB/LSS data implies that the DM-baryon
interaction has to be quite weak, a statement we quantify
precisely below.

CMB/LSS constraints to the baryon-DM interaction have been
obtained in a variety of previous papers.  The first such paper
\cite{Chen:2002yh} considered velocity-independent scattering
and obtained limits from 2dFGRS LSS data and from CMB data from
the set of
suborbital missions that preceded the Wilkinson Microwave
Anisotropy Probe (WMAP).  Ref.~\cite{Sigurdson:2004zp}
considered constraints for DM particles that interact with
baryons through a DM electromagnetic dipole using early WMAP
data.  Ref.~\cite{Boehm:2004th} considered constraints to the
baryon-DM interaction from the damping of small-scale structure.  
DM millicharge have been discussed in \cite{Dolgov:2013una,McDermott:2010pa,Dubovsky:2001tr,Dubovsky:2003yn,Davidson:2000hf}.  Constraints to the
baryon-DM interaction from galaxy clusters were presented in
Refs.~\cite{Chuzhoy:2004bc,Hu:2007ai}.  There is also a
 body of work on direct detection
\cite{Starkman:1990nj,Erickcek:2007jv,Albuquerque:2003ei,Sanglard:2005we,Akerib:2005kh,Akerib:2005za}, 
gravitational lensing \cite{Natarajan:2002cw}, gas temperature in clusters \cite{Qin:2001hh}, big-bang
nucleosynthesis \cite{Mohapatra:1998nd}, cosmic rays \cite{Cyburt:2002uw,Wandelt:2000ad}, and cosmic gamma rays
\cite{Mack:2012ju}. If DM annihilates, then constraints on the
Earth's heat flow \cite{Mack:2007xj,Mack:2012ju} become
important, though this constraint disappears if annihilation is
absent as could be the case for asymmetric dark matter
\cite{Kaplan:2009ag,Zurek:2013wia}. Related studies were addressed in the context of
strongly self-interacting dark matter particles, first suggested by Ref. \cite{Spergel:1999mh}.

Our work extends and improves upon previous work on CMB/LSS
constraints to the baryon-DM interaction in two ways.
First of all, we provide a model-independent analysis,
considering cross sections that scale with DM-baryon relative
velocity $v$ as $v^n$ with arbitrary power-law index $n$.
Second, we work out from first principles the effect of scattering in cosmological perturbation theory, highlighting the interplay between bulk velocities and thermal velocities. In prior work it was assumed that the relative 
baryon-DM bulk velocities are small compared to the thermal
velocities.  We show that this assumption becomes invalid at
redshifts $z\lesssim 10^4$, signaling the breakdown of the ordinary linear theory.  We  introduce an approximation
to account for this nonlinearity, and estimate the 
theoretical uncertainty.  We use for our numerical results what
we believe to be a conservative estimate of the magnitude of the
effects of this nonlinearity.  We then use not only the recent
Planck data, but also include for the first time in this context constraints to
the matter power spectrum from the Lyman-alpha forest measurements.
Inclusion of Lyman-alpha data improves, as we will see, upon
constraints obtained from the CMB by several orders of
magnitude.

The main result of this paper is that a baryon-DM interaction
strong enough to affect the global structure of a galaxy like our own Milky Way, through scattering at low redshift, is excluded. As we show, this result is
model-independent, and cannot be circumvented by plausible
particle-physics model building, as long as the DM we infer
locally by galactic rotation curves and cluster dynamics is the
same DM that affects the linear collapse of density
perturbations in the early Universe. 

This paper is organized as follows.  In
\S\ref{sec:general_evolution_eqs} we compute the drag force
produced by the baryons on the dark-matter fluid due to the
DM-baryon interactions.  In \S\ref{sec:perturbations} we derive
the modified Boltzmann equations for the dark matter and the
baryons, and provide simple analytical estimates of our results
before presenting our numerical constraints in
\S\ref{sec:CMB_constraints}.  In \S\ref{sec:satellite_galaxies}
we show the effect of DM-baryon interactions on the halo mass
function. Finally, we present our conclusions in
\S\ref{sec:discussion}.  We derive the heating rate in Appendix
\ref{sec:thermalization} and give the DM-baryon momentum transfer rate beyond leading order in bulk velocities in Appendix \ref{sec:Gn}. In Appendix \ref{app:models} we compare our results to previous studies focusing on specific particle-physics models, including velocity-independent cross section as well as dark matter millicharge and electromagnetic dipole moments. 

\section{The baryon--dark-matter drag force}
\label{sec:general_evolution_eqs}

Here we calculate the drag force $d\vec{v}_\chi/dt$ per unit
mass exerted by the baryons on the dark-matter fluid as a
consequence of the DM-baryon interaction.  Our calculations are
valid for $z \lesssim 10^9$, where DM particles of mass $m_\chi
\gtrsim\rm MeV$ are non-relativistic, and we assume that DM
particles and baryons are non-relativistic throughout our
analysis.

Consider a DM particle of velocity $\vec v_\chi$, moving in a
background of non-relativistic baryons with thermal velocity
distribution $f_b(v_b)$ as a function of baryon velocity $v_b$,
in a frame where the baryon distribution is isotropic.  In this
frame, the baryon velocity distribution depends only on the
magnitude $v_b$ of the velocity, not on its direction, and
momentum exchange with baryons drives the DM velocity towards
zero. The change in DM momentum per collision is
\ba
     \Delta \vec p_\chi=\frac{m_\chi m_b}{m_\chi+m_b}\left|\vec
     v_\chi-\vec v_b\right|\left(\hat n-\frac{\vec v_\chi-\vec
     v_b}{\left|\vec v_\chi-\vec v_b\right|}\right),
\ea
to leading order in velocities, where $\hat n$ is the direction of
the scattered DM particle in the center-of-mass frame and $m_b$
and $m_\chi$ respectively the baryon and dark-matter masses.
The acceleration experienced by the DM is then
\ba
     \frac {d\vec v_\chi}{dt} &=&\frac{\rho_b}{m_\chi+m_b}\int
     dv_bv_b^2f_b(v_b)\int\frac{d\hat n_b}{4\pi}\int d\hat
     n\left(\frac{d\sigma\left(|\vec v_\chi-\vec
     v_b|\right)}{d\hat n}\right)|\vec v_\chi-\vec
     v_b|^2\left(\hat n-\frac{\vec v_\chi-\vec v_b}{|\vec
     v_\chi-\vec v_b|}\right)\\
     &=&-\frac{\rho_b\,\vec
     v_\chi}{m_\chi+m_b}\frac{v_\chi^4}{2}\int_0^\infty
     dxx^2f_b(xv_\chi)\int_{-1}^1
     dy\,\bar\sigma\left(v_\chi\sqrt{1+x^2-2xy}\right)
     \sqrt{1+x^2-2xy}\left(1-xy\right),\no
\ea
where $\rho_b$ is the baryon mass density.
Here $d\sigma(v)/d\hat n$ is the differential cross section for
baryon-DM scattering as a function of the baryon-DM relative
velocity $v$.  We also define the momentum-transfer cross section,
\ba
\label{eq:barsig}
     \bar\sigma(v)\equiv\int
     dc_\theta(1-c_\theta)\left(\frac{d\sigma(v)}{dc_\theta}\right),
\ea
where $\theta$ is the center-of-mass scattering angle,
$c_\theta=\cos\theta$.  

We take the cross section
$\bar\sigma(v)$ to have a power-law dependence on baryon-DM
relative velocity $v$ (where the speed of light is $c=1$),
\ba
\label{eq:sign} 
     \bar\sigma(v)=\sigma_0 v^n.
\ea
We work out the momentum exchange for arbitrary $n$, and present
results for selected values. 
Note that, e.g., $n=-1$ comes about from a Yukawa potential
(massive-boson exchange), $n=-2$ occurs if DM has an electric
dipole moment \cite{Sigurdson:2004zp}, and $n=-4$ occurs for DM
millicharge \cite{Dolgov:2013una,McDermott:2010pa,Dubovsky:2001tr,Dubovsky:2003yn}.  We take $\bar\sigma$ here to be
the scattering
cross section for DM from hydrogen, and then correct it below to
account for the additional scattering from helium.

In the early Universe, the baryon velocity distribution (in the
isotropic frame) is
\ba 
     f_b(v_b)=\sqrt{\frac{2}{\pi}}\frac{1}{u_b^3}e^{-(v_b/u_b)^2/2},
\ea
with $u_b^2=T_b/m_b$.  For the DM we assume a similar Maxwell 
distribution, 
\ba 
     f_\chi(v_\chi)=\sqrt{\frac{2}{\pi}}\frac{1}{u_\chi^3}
     \mathrm{exp}\left[-\frac{\left(\hat
     n_\chi v_\chi-\vec
     V_\chi\right)^2}{2u_\chi^2} \right],
\ea 
with $ u_\chi^2=T_\chi/m_\chi$,
boosted with peculiar velocity $\vec V_\chi$ with respect to
the baryon frame.  The rate of change of the peculiar velocity
is\footnote{The subscript $\chi$ on $d\vec V_\chi/dt$ is there
to emphasize that the deceleration of the DM fluid is different
than that of the baryons, due to the difference in inertia of
the two fluids, even though the instantaneous relative velocity
is of course the same as measured in either the isotropic DM or
baryon frames.}
\ba
     \frac{d\vec V_\chi}{dt}&=&\int\frac{d\hat n_\chi}{4\pi}\int
     dv_\chi v_\chi^2f_\chi(v_\chi)\frac{d\vec v_\chi}{dt}.
\ea

In general, there are two velocity scales that enter $d\vec
V_\chi/dt$. The first is the thermal velocity dispersion,
\be
     \dispersion
     =\VEV{\left(\vec v_\chi-\vec
     v_b\right)^2} =
     3\left(\frac{T_b}{m_b}+\frac{T_\chi}{m_\chi}\right),
\ee
where $\langle...\rangle$ denotes thermal average. 
The second is the peculiar velocity $V_\chi$ itself. In the limit
where the peculiar velocity is smaller than the
velocity dispersion, $V_\chi^2 < \dispersion$, we find
\ba
\label{eq:VdotTherm}
     \frac{d\vec V_\chi}{dt}&=&-\vec
     V_\chi\,\frac{c_n \rho_b\sigma_0
     \left(\frac{\dispersion}{3}\right)^{\frac{n+1}{2}}}
     {m_\chi+m_b},
\ea
at leading order in $\left(V_\chi^2/\dispersion\right)$, 
with
\be
     \label{eq:Rsc}
     c_n=\frac{2^{\frac{n+5}{2}}\Gamma
     \left(3+\frac{n}{2}\right)}{3\sqrt{\pi}},
\ee
evaluating to 
$c_n\approx\{0.27, 0.33, 0.53, 1, 2.1, 5, 13,
35, 102\}$ for 
$n=\{-4,-3,-2,-1,0,1,2,3,4\}$.

In the limit that the peculiar velocity is larger than the
velocity dispersion, the calculation reduces to the deceleration
of the relative motion between two cold flows. The deceleration
of the DM fluid in this case is given by
\ba
\label{eq:VdotPec}
     \frac{d\vec V_\chi}{dt}&=&-\vec
     V_\chi\,\frac{\rho_b\,\sigma_0\left|V_\chi
     \right|^{n+1}}{m_\chi+m_b}
     ,
\ea
at leading order in $\left(\dispersion/V_\chi^2\right)$. 

Note that in general, the dependence of $dV_\chi/dt$,
the drag force per unit mass, on the baryon-DM relative velocity
is not linear.  In the limit 
$V_\chi^2 \ll \dispersion$, 
the dependence reduces to linear.  In the
opposite limit, 
$V_\chi^2 \gg \left\langle \Delta\vec
v^2\right\rangle$, 
the dependence on $V_\chi$ is nonlinear
unless $n=-1$.  In the early Universe, as we look further
backwards in time, there comes a time when typical peculiar
velocities become small in comparison to the thermal velocity
dispersion. The transition occurs around redshift $z\sim10^4$
(see Fig.~\ref{fig:vmtherm}). At earlier times (higher
redshift), Eq.~(\ref{eq:VdotTherm}) then tell us that we may use
linear perturbation theory in order to calculate the evolution
of the peculiar velocity $\vec V_\chi$. In what follows, we use
this observation to calculate precisely the evolution of modes
at high redshift in order to compare with cosmological data. We
discuss later on the complication arising at $z<10^4$, where the
problem becomes nonlinear.


\section{Linear cosmological perturbations with
     dark-matter--baryon interactions}
\label{sec:perturbations}

\subsection{Boltzmann equations}
\label{sec:evolution_eqs}

We now consider the modifications to the Boltzmann equations for
dark matter and baryons that arise from the baryon-DM coupling.
We work in synchronous gauge, following the notation and
conventions of Ref.~\cite{Ma:1995ey}.  We allow for a nonzero
peculiar velocity for DM that arises from the interaction with
baryons~\cite{Chen:2002yh,Sigurdson:2004zp} and defined so that
the DM peculiar velocity vanishes in the absence of scattering.
The evolution equations for the DM and baryon density
fluctuations, $\delta_\chi$ and $\delta_b$ respectively, and
velocity divergence, $\theta_\chi$ and $\theta_b$, respectively,
are given for a Fourier mode of wavenumber $k$ by
\ba
     \dot{\delta_\chi}&=& -\theta_\chi-{\dot{h}\over 2}, \qquad
     \dot{\delta_b}= -\theta_b-{\dot{h}\over 2},\nonumber\\
     \dot{\theta_\chi}&=&-{\dot{a}\over a}\theta_\chi +
     c_\chi^2k^2\delta_\chi +
     R_\chi\left(\theta_b-\theta_\chi\right),\nonumber\\
     \dot{\theta_b}&=&-{\dot{a}\over a}\theta_b +
     c_b^2k^2\delta_b+
     R_\gamma\left(\theta_\gamma-\theta_b\right)
     +\frac{\rho_\chi}{\rho_b}R_\chi
     \left(\theta_\chi-\theta_b\right),\nn\\
\label{eq:boltzmann}
\ea
where $\rho_\chi$ ($\rho_b$) is the DM (baryon) mass density, and an
overdot denotes derivative with respect to conformal
time.  We derive the DM-baryon momentum-exchange coefficient
$R_\chi$ below in Sec.~\ref{ssec:Rs}.

The DM and baryon temperatures evolve according to
\ba
     \label{eq:dotT}
     \dot T_\chi&=&-2{\dot{a}\over a}T_\chi +
     \frac{2\,m_\chi}{m_\chi+m_{\rm H}}\,R'_\chi \,
     \left(T_b-T_\chi\right),\no\\
     \dot T_b&=&-2{\dot{a}\over a} T_b+
     \frac{2\,\mu_b}{m_\chi+m_{\rm H}} \, \frac{\rho_\chi}{\rho_b} \,
     R'_\chi\, \left(T_\chi-T_b\right) \nonumber \\
     & & + \frac{2\mu_b}{m_e}R_{\gamma} \left(T_\gamma-T_b\right).
\ea
Here, $\mu_b\simeq
m_{\rm H}\left(n_H+4n_{\rm He}\right)/\left(n_H+n_{\rm He}+n_e\right)$ is the
mean molecular weight for the baryons, and
$R_{\gamma}=(4/3)(\rho_\gamma/\rho_b) a n_e\sigma_T$ is the
usual Compton collision term \cite{Ma:1995ey}. The thermalization rate
$R'_\chi$ is related to the momentum exchange rate $R_\chi$
(with $R'_\chi\to R_\chi$ in the heavy DM limit) and is given in
Sec.~\ref{ssec:Rs} below.

Our calculations apply to cold DM with mass $m_\chi>\rm
MeV$, that is non-relativistic at redshift $z<10^9$.  We
therefore neglect possible direct momentum transfer between the
photon and DM fluids, and consider only direct interaction with
baryons.  For the calculations we will be interested in, the DM
sound speed $c_\chi^2$ is unimportant, and we neglect
the corresponding term in what follows.

\subsection{The momentum-exchange rate coefficient}\label{ssec:Rs}

If the peculiar velocity is small compared with the thermal
velocity---i.e., if $V_\chi^2 \ll \dispersion$---then the
DM-baryon momentum-exchange and thermalization rate coefficients, appearing in Eqs.~(\ref{eq:boltzmann}) and~(\ref{eq:dotT}), can be read from Eqs.~(\ref{eq:VdotTherm}) and~(\ref{eq:dQdt}) to be
\ba
\label{eq:Rs}
     R_\chi&=&\frac{a\,c_n\,\rho_b\,\sigma_0}{m_\chi+m_{\rm H}}
     \left(\frac{T_b}{m_{\rm H}}+\frac{T_\chi}{m_\chi}\right) ^
     {\frac{n+1}{2}}\mathcal{F}_{\rm He}
\ea 
and
\be 
     R'_\chi = R_\chi\left[1+\frac{3m_{\rm H}}{m_\chi+4m_{\rm H}}
     \left(\frac{1-f_{\rm He}}{\mathcal{F}_{\rm He}} -
     1\right)\right],
\ee
respectively, with $R'_\chi\simeq R_\chi$ for heavy DM.  

We include a correction factor,
\ba\label{eq:fHe}
     \mathcal{F}_{\rm He}&=&1-f_{\rm He}+f_{\rm He}
     \frac{\sigma_{\rm He}}{\sigma_0}
     \frac{1+\frac{m_{\rm H}}{m_\chi}}{1+\frac{4m_{\rm H}}{m_\chi}}
     \left(\frac{1+\frac{T_\chi\,m_{\rm H}}{T_b\,m_\chi}}
     {1+\frac{4T_\chi\,m_{\rm H}}{T_b\,m_\chi}}
     \right)^{\frac{n+1}{2}}\nonumber \\
     &\simeq & 1+0.24 \,
     \left(\frac{\sigma_{\rm He}}{\sigma_0}-1\right),
\ea
for scattering from helium with mass $m_{\rm He}\simeq 4 m_{\rm H}$ and
mass fraction $f_{\rm He}\simeq 0.24$.  The approximation on the
second line of Eq.~(\ref{eq:fHe}) is applicable if the DM is heavier than helium.
The value of $\mathcal{F}_{\rm He}$ depends on the
ratio $\left(\sigma_{\rm He}/\sigma_0\right)$ between the
cross section for scattering on helium to that for scattering on
hydrogen. Plausible numerical values are, e.g.,
$\mathcal{F}_{\rm He}=4.6$ or $\mathcal{F}_{\rm He}=1.7$, valid for DM mass above a few GeV
with the same amplitude for
scattering from protons and neutrons and, respectively, coherent or incoherent scattering on helium. Nevertheless, as $\mathcal{F}_{\rm He}$ involves some model dependence, in reporting our numerical results we conservatively set $\left(\sigma_{\rm He}/\sigma_0\right)=0$, fixing $\mathcal{F}_{\rm He}=0.76$ unless explicitly stated otherwise. 

For $V_\chi^2 \ll \dispersion$, the coefficient $R_\chi$ is {\it in}dependent
of $\theta_\chi-\theta_b$, and the DM-baryon drag that appears
in Eq.~(\ref{eq:boltzmann}) is linear in the velocity perturbation.  The usual linear-theory
approach, obtained by solving the linearized Boltzmann
equations independently for each Fourier mode, is valid.

\begin{figure}[ht!]
\centering
\includegraphics[width=0.47\textwidth]{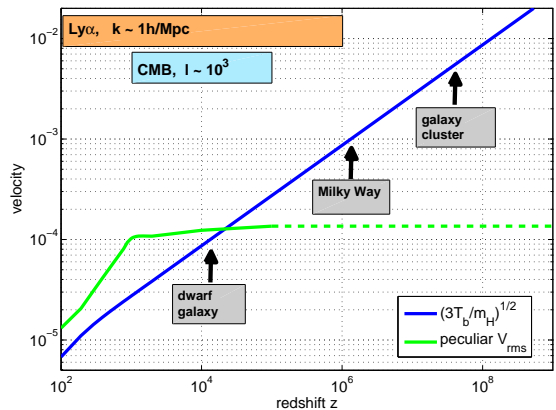}
\caption{Cosmological proton thermal velocity (blue), and
     peculiar baryon-DM relative velocity (green). The redshifts
     probed by CMB and Lyman-$\alpha$ forest measurements are
     roughly marked by blue and orange boxes,
     respectively.}
\label{fig:vmtherm}
\end{figure}

\begin{figure}[ht!]
\centering
\includegraphics[width=0.47\textwidth]{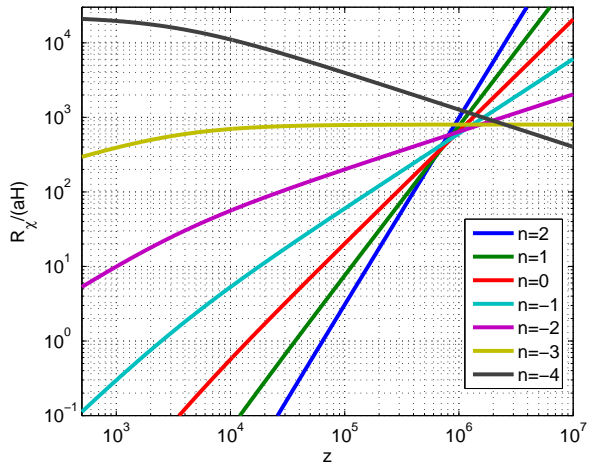}
\caption{Momentum-transfer rate $R_\chi$
     vs.\ redshift for different values of $n$ between -4 to
     +2. All of the curves are normalized to satisfy a mean free
     path of $\sim0.5$~Mpc for proton scattering on DM at the MW
     solar cycle; see Eq.~(\ref{eq:SS}) and the discussion
     around it.}
\label{fig:rates}
\end{figure}

However, this assumption $\left(V_\chi^2 \ll \dispersion\right)$ is not
always valid.  The rms DM-baryon relative velocity is given
by~\cite{Tseliakhovich:2010bj}
\ba
     V_{\rm RMS}^2 & = &\left\langle \vec V_\chi^2\right\rangle_\xi =\int\frac{dk}{k} \Delta_\xi
     \left(\frac{\theta_b-\theta_c}{k}\right)^2,
\ea 
where $\langle ...\rangle_\xi$ denotes an average with respect to the primordial curvature perturbation and $\Delta_\xi\simeq2.4\times10^{-9}$ is the primordial
curvature variance per log $k$.  The value of $V_{\rm RMS}$ is shown as the
green curve in Fig.~\ref{fig:vmtherm} (for $z>10^5$, we replace the direct calculation of $V_{\rm RMS}$ by analytic estimate). The
peculiar velocity becomes larger than the baryon thermal
velocity (the blue curve) below $z\sim10^4$. At later times
(lower redshift), the effect of baryon-DM scattering will have a
nonlinear dependence on peculiar velocity, as discussed at the end of
Section~\ref{sec:general_evolution_eqs}.  This implies that the
drag terms in the Boltzmann equations for $\theta_b$ and
$\theta_c$ are no longer linear in $\theta_b$ and $\theta_c$, mixing together the evolution of
different Fourier modes.

To improve the domain of validity of linear theory, we extend the
rate coefficient of Eq.~(\ref{eq:Rs}) by summing together the thermal and peculiar rms velocity dispersion,
\ba
\label{eq:Rstotal}
      R_\chi&\to&\frac{a\rho_b\sigma_0 \mathcal{F}_{\rm
     He}}{m_\chi+m_{\rm H}} c_n\left( \frac{T_b}{m_{\rm H}}+\frac{T_\chi}{m_\chi}+\frac{V_{\rm RMS}^2}{3}
     \right)^{n+1\over 2}.
\ea 
This ``mean-field'' approach is generally valid at $z>10^4$, but should also apply at later times for modes with wavelengths $k\lesssim0.1$~Mpc$^{-1}$, long compared with those that contribute most to the rms peculiar velocity at $z<10^4$.
It also obtains the correct parametric scaling for short-wavelength modes at low redshift, though the numerical coefficient $c_n$, encapsulating thermal velocity integrals, needs to be modified in this limit. 

As we show below, models with $n\geq-2$ are strongly constrained by LSS data for which the relevant dynamics occurs at relatively high redshift, $z\gg10^4$, where Eq.~(\ref{eq:Rs}) (and similarly, in this regime, Eq.~(\ref{eq:Rstotal})) is reliable. For all such models ($n\geq-2$), our constraints derived using Eq.~(\ref{eq:Rstotal}) are directly applicable. Models with scattering cross section that increases rapidly at decreasing velocity, $n\leq-3$, are less constrained by LSS and, instead, more strongly constrained by CMB data, that is sensitive to perturbation evolution at $z<10^4$. For such models ($n\leq-3$), using Eq.~(\ref{eq:Rstotal}) rather than Eq.~(\ref{eq:Rs}) makes a significant difference. This means that our linear calculation is less reliable, and that nonlinear (and thus non-gaussian) effects cannot be neglected. Nevertheless, note from Eq.~(\ref{eq:Rsc}) that for models with $n\leq-3$ the coefficient $c_n$ is smaller than unity. Comparing to Eq.~(\ref{eq:VdotPec}), describing the momentum transfer in the cold-flow limit, we learn that our use of   Eq.~(\ref{eq:Rstotal}) is likely conservative, meaning that a more detailed treatment of the nonlinear $z<10^4$ regime for these models would most likely yield even stronger constraints. 

We conclude that for our practical purpose of obtaining conservative limits on the DM-baryon interactions, Eq.~(\ref{eq:Rstotal}) is adequate for deriving model-independent results, and we use it in the bulk of our analysis. For completeness, when quoting our numerical results (see Sec. \ref{sec:CMB_constraints}), we also report results using Eq.~(\ref{eq:Rs}) instead of (\ref{eq:Rstotal}). 
We leave a precise treatment of the nonlinear effects induced by
DM-baryon couplings, as well as the resulting secondary
non-Gaussianity, for future work.

In our numerical analysis, we modify CAMB to include the new
perturbation equations above. In what follows we describe the
analytic behavior of the solution before moving on to the
results.

\subsection{Analytic discussion}

Before moving to the numerical results, we provide some simple
analytic estimates. We focus on redshifts $z>300$ when
the baryon temperature follows the CMB temperature, $T_b\simeq
T_\gamma$, due to Thomson scattering. Since the dark matter does
not scatter from photons, we expect that  $T_\chi<T_b$.
Then, if $m_\chi > m_b$, we can for simple estimates
neglect $T_\chi/m_\chi$ relative to $T_b/m_b$.
Comparing the momentum exchange rate $R_\chi$ to the comoving Hubble expansion rate
$aH=(\dot a/a)$, we have
\ba
     \label{eq:approx} 
     \frac{R_\chi}{aH} &= & \frac{c_n\,\rho_b\,
     \sigma_0}{H\,m_\chi}
     \left(\frac{T_b}{m_{\rm H}}\right)^{\frac{n+1}{2}}
     \mathcal{F}_{\rm He}\no\\
     &\simeq&10 \,c_n\, \frac{\sigma_0/m_\chi}{\rm cm^2/g} \,
     \frac{H\left(z=10^5\right)}
     {H(z)}\,\left(\frac{z}{10^5}\right)^{3.5} \nonumber \\
     & & \times
     \left[1.6\cdot10^{-4}
     \left(\frac{z}{10^5}\right)^{\frac{1}{2}} \right]^{n}
     \mathcal{F}_{\rm He}.
\ea

At high redshift $z>10^4$, Eq.~(\ref{eq:approx}) can be directly translated to a model-independent constraint on the interaction between DM and baryons in a contemporary ($z=0$) system like our own Milky Way (MW) galaxy. To see this, note that if we choose $\sigma_0$ for the different values of $n$ 
so that $\bar\sigma(v)$ is the same at the velocity $v_{\rm MW}=10^{-3}$,
characteristic of the virial velocity in a MW-type
halo, then the scattering rates for all $n$ will coincide, up to
the $\mathcal{O}(1)$ coefficients $c_n$ of Eq.~(\ref{eq:Rsc}), at a common
redshift $z$ where the thermal velocity is approximately equal to $v_{\rm MW}$,
as in Fig.~\ref{fig:vmtherm}.  We illustrate
this behavior in Fig.~\ref{fig:rates}, where we normalize
$\sigma_0$ so that the scattering rate for all $n$ will yield a baryon-DM mean
free path of about 1~Mpc in our MW galaxy. 
We learn that with this normalization, for all values of $n$, the rate of momentum
exchange between DM and baryons is much faster than expansion
for $z>10^5$.  Comparing to the reach of CMB and Lyman-$\alpha$
observables, which are sensitive to the evolution of linear
perturbations at these redshifts, we expect that linear
cosmology places strong
constraints on baryon-DM scattering for any velocity dependence,
implying mean free path $\lambda\gg1$~Mpc, orders of magnitude
larger than the $\mathcal{O}(10\,{\rm kpc})$ scale size of the
galaxy itself.

\section{Numerical results: CMB and Lyman-alpha constraints}
\label{sec:CMB_constraints}

We incorporate the parameter $\sigma_0$ into a Markov Chain
Monte Carlo (MCMC) likelihood analysis\footnote{http://cosmologist.info/cosmomc/}~
\cite{Lewis:2002ah}
of {\it Planck} data \cite{Planck:2013kta} and measurements of the Lyman-$\alpha$ flux
power spectrum from the {\it Sloan Digital Sky Survey}
\cite{McDonald:2004xn}. We checked that adding ACT \cite{Dunkley:2013vu} and SPT \cite{Hou:2012xq} data to the CMB analysis makes only a small improvement to the results. We run the MCMC to determine 95\% confidence level (CL)
constraints on $\sigma_0$, fixing the value of $m_\chi$ and of
the power-law index $n$ in each run.  

Having obtained a
constraint on $\sigma_0$ in this way for $m_\chi=10$~GeV, we present our result as a constraint on
$(\sigma_0/m_\chi)$, valid for any value of $m_\chi$ subject to $m_\chi \gg m_{\rm H}$, and quoted separately for different 
values of $n$. Note that, in the limit of $m_\chi\gg m_{\rm H}$,
there is no dependence on $n$ in the scaling of the bound as
function of $m_\chi$ for fixed $\sigma_0$, to leading order in
$(m_{\rm H}/m_\chi)$. This is so because all the dynamical
difference between the models is contained in the velocity
dependence, where the thermal dispersion becomes dominated by
the baryons, $\dispersion\approx3\left(T_b/m_b\right)$ to
leading order in $\left(m_{\rm H}/m_\chi\right)$. While we do
not discuss here in detail the limit $m_\chi<m_{\rm H}$, we note
that the set of equations presented in
Sec.~\ref{sec:perturbations} provides all of the information
required to evaluate the bounds in the low-$m_\chi$
limit, as long as the DM is non-relativistic throughout the time
of interest $z\lesssim10^9$ (satisfied for
$m_\chi\gtrsim1$~MeV).

We determine joint constraints on $\sigma_0$ and the basic set of 
$\Lambda$CDM cosmological parameters,
\be
     p_\mu=\{\Omega_bh^2,\Omega_\chi
     h^2,\tau,\theta,A_s,n_s\}.
\label{eq:parameters}
\ee
Here $\Omega_bh^2$ is the physical baryon density, $\Omega_\chi
h^2$ is the physical dark matter density, $\tau$ is the
reionization optical depth, and $\theta$ is the angular size of
the sound horizon at recombination. We ignore tensor modes and
assume a flat geometry.

Our numerical results are summarized in Table~\ref{tab:1}. In
obtaining these bounds, instead of solving for $T_\chi$ [which
can easily be done using Eq.~(\ref{eq:dotT})] we simply set
$T_\chi=T_b$. The induced error is of
$\mathcal{O}\left(m_{\rm H}/m_\chi\right)$ for heavy DM. 

\begin{table*}[htdp]
\begin{center}
\begin{tabular}{||c||c|c|c||}
\hline
$n$ & CMB (95\%CL, cm$^2$/g) & CMB + Lyman-$\alpha$ (95\%CL, cm$^2$/g) & $\lambda$ (MW)\\
\hline\hline
-4&$1.8\times10^{-17}$&$1.7\times10^{-17}$&$27$ Gpc\\
\hline
-2&$3.0\times10^{-9}$&$6.2\times10^{-10}$&$738$ Mpc\\
\hline
-1&$1.6\times10^{-5}$&$1.4\times10^{-6}$&313~Mpc\\
\hline
0&0.12&$3.3\times10^{-3}$&138~Mpc\\
\hline
+2&$1.3\times10^5$&$9.5\times10^3$&46~Mpc\\
\hline
\end{tabular}
\caption{95\%CL constraints on $\left(\sigma_0/m_\chi\right)$
     from CMB alone (with Planck data) and from CMB in combination with
     Lyman-$\alpha$ data from the SDSS. Results are valid for $m_\chi\gg m_{\rm H}$, and
     conservatively neglect scattering from helium, setting $\mathcal{F}_{\rm He}=0.76$ (adding coherent isospin-independent scattering on helium would tighten the bounds by a factor of 6). First column: power-law
     index $n$ of Eq.~(\ref{eq:sign}). Second column: CMB alone,
     constraint in units of cm$^2$/g. Third column: combined CMB
     and Lyman-$\alpha$. Fourth column: minimal mean free path
     for baryon scattering on DM in the MW solar cycle
     ($\rho_\chi\sim0.4$~GeV/cm$^3$, $v=v_{\rm MW}\sim10^{-3}$), using the
     CMB + Lyman-$\alpha$ constraint.}\label{tab:1}
\end{center}
\label{default}
\end{table*}%

These constraints are obtained using the momentum-transfer rate
given in Eq.~(\ref{eq:Rstotal}). As discussed at the end of
Sec.~\ref{ssec:Rs}, at redshift $z<10^4$ Eq.~(\ref{eq:Rstotal})
provides only an approximate treatment of the perturbation
equations as the full evolution becomes nonlinear\footnote{This
issue is relevant for models with $n\neq-1$. For $n=-1$,
Eqs.~(\ref{eq:Rs}) and~(\ref{eq:Rstotal}) coincide.}. To estimate the impact of our approximation, we compare
the constraints reported in Tab.~\ref{tab:1} to the constraints
obtained using Eq. (\ref{eq:Rs}), instead
of~(\ref{eq:Rstotal}). For the the $n=-2$, $n=0$, and $n=+2$
models, we find that the CMB+Lyman-$\alpha$ constraints exhibit
essentially no change. This happens because for these models,
Lyman-$\alpha$ dominates the constraint, and the matter power
spectrum on the scales probed by Lyman-$\alpha$ is determined by
mode evolution at $z\gg10^4$, where Eqs.~(\ref{eq:Rs})
and~(\ref{eq:Rstotal}) are equally valid. In contrast, the model
with $n=-4$ is constrained primarily by the CMB data, and is
sensitive to the appearance of $V_{\rm RMS}^2$ in
Eq.~(\ref{eq:Rstotal}) that regularizes an otherwise decreasing
thermal velocity. Using Eq.~(\ref{eq:Rs}) instead
of~(\ref{eq:Rstotal}) for the model with $n=-4$, we would find
an artificially stronger bound,
$\left(\sigma_0/m_\chi\right)<1.4\times10^{-18}$, more
constraining by a factor of 10 compared with the number we quote
in Tab.~\ref{tab:1}. We believe that our simplified analysis of
the $n=-4$ case in the nonlinear regime is conservative, and
leaves room for significant improvement of the constraints. This
could be of particular interest as $n=-4$ arises in simple
particle physics models where DM has a small electric charge. 

\begin{figure}[ht!]
\centering
\includegraphics[width=0.45\textwidth]{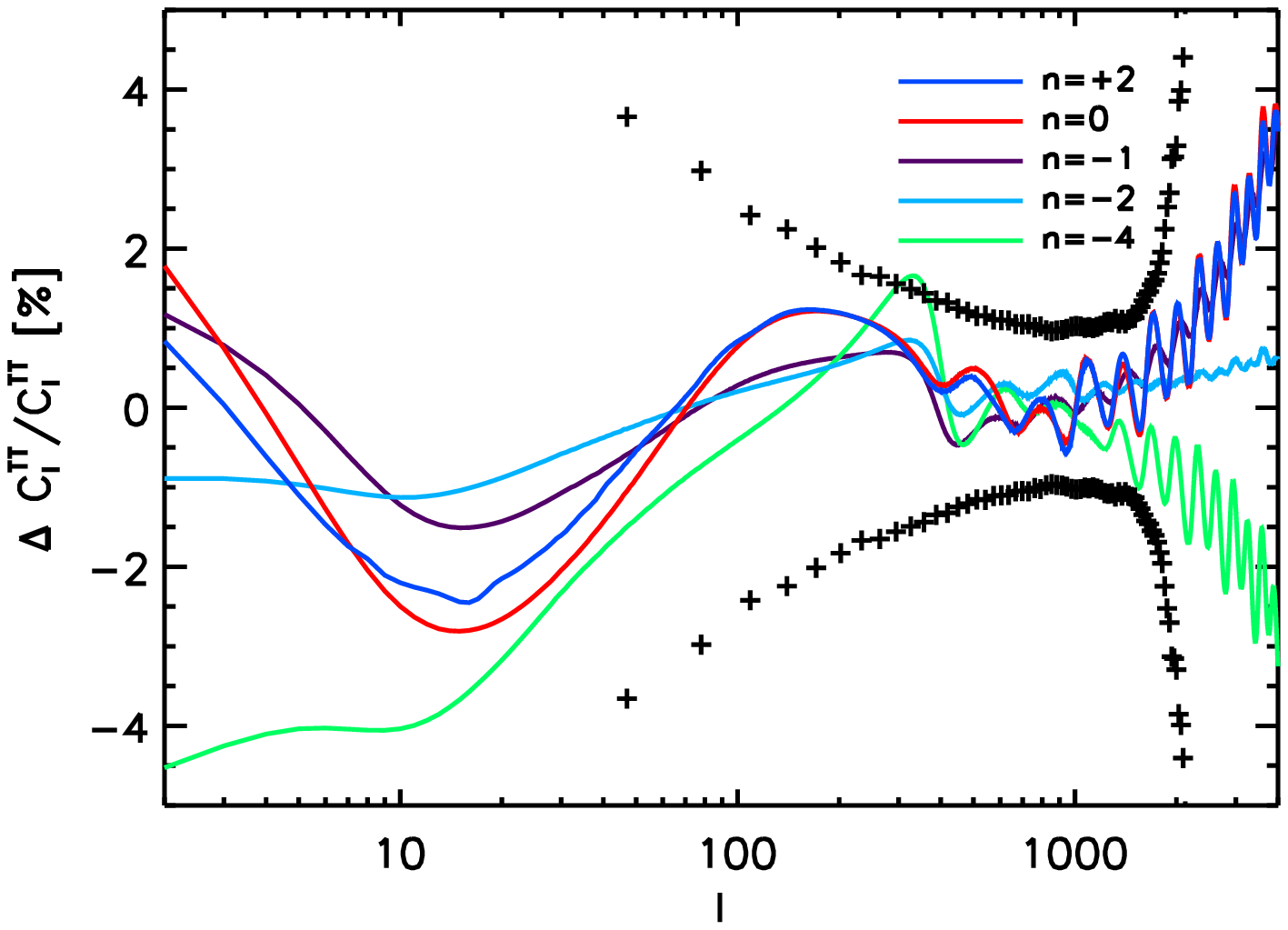}
\includegraphics[width=0.45\textwidth]{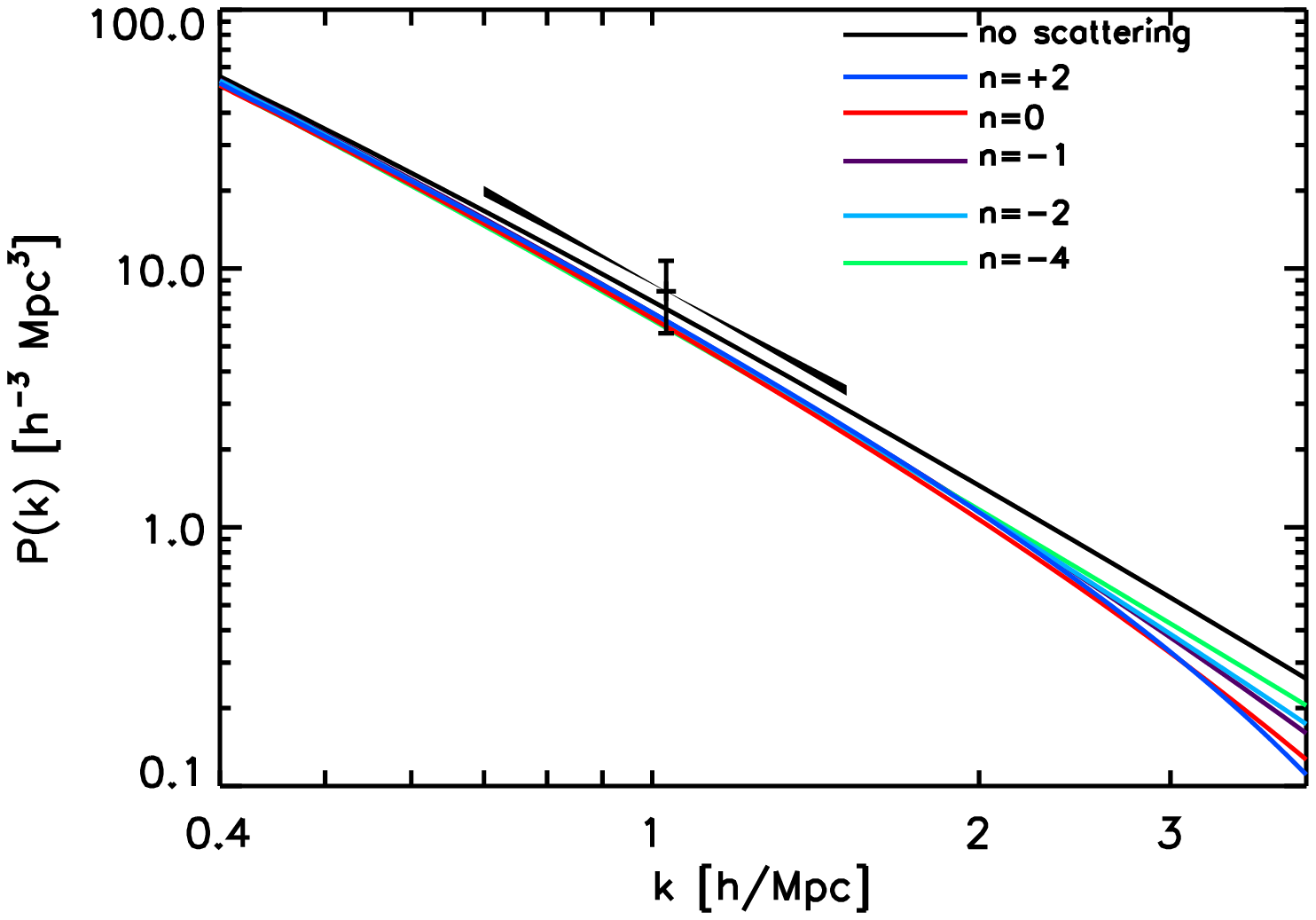}
\caption{Left panel: Relative difference of the CMB power
     spectra of models with  different velocity-dependent cross sections to the best fit $\Lambda$CDM model. The cross sections of the different models correspond to the 95\%CL limit from the CMB $+$
     Lyman-$\alpha$ analysis (see Tab.~\ref{tab:1}), while all other cosmological parameters are taken to optimize the likelihood for this given cross section. Right panel: Matter power spectra
     at $z=3$. The data point corresponds to the linear theory best fit amplitude using Lyman-$\alpha$ data from \cite{McDonald:2004xn}. The error bar corresponds to the $95\%$ CL limit on the amplitude. The black band denotes the range of linear matter power spectra slopes allowed at the $95\%$ CL limit at $k=1.03$ h/Mpc.}
\label{fig:ClPkdiff}
\end{figure}

In Fig.~\ref{fig:ClPkdiff} we show the effect of DM-baryon
scattering on the CMB and matter power spectra, using for the plots the 95\% CL limits
from the CMB $+$ Lyman-$\alpha$ chains, taken from Tab.~\ref{tab:1}. We
add in Fig.~\ref{fig:ClPkdiff} (right panel) the experimental Lyman-$\alpha$ data point used in the likelihood analysis, at $k=1.03$ h/Mpc, showing the $95 \%$ CL limit of both amplitude and slope. In the CMB plot, we denote the $\pm1\sigma$ error bars of Planck, including beam noise and cosmic variance, as black $(+)$ marks. 

\begin{figure}[ht!]
\centering
\includegraphics[width=0.45\textwidth]{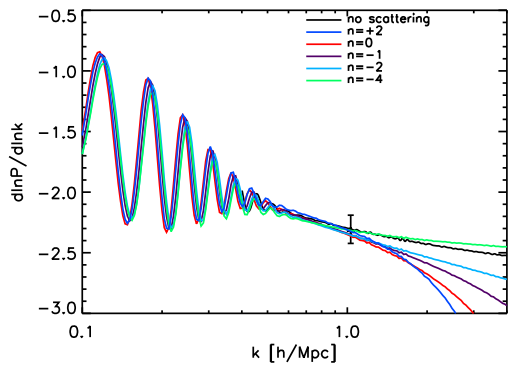}
\caption{Slope $d\ln{P}/d\ln{k}$ for the different models, as a function of wave number. The data point corresponds to the best fit value of the linear matter power spectrum slope from the Lyman-$\alpha$ measurement in Ref. \cite{McDonald:2004xn}, and the error bar on the point corresponds to the $95\%$ CL limit.}
\label{fig:slope_at95CL_CMB_Lya_all_models}
\end{figure}

In Fig. \ref{fig:slope_at95CL_CMB_Lya_all_models} we show separately the slope of the linear matter power spectra for the different models, along with the experimental value and its $95\%$CL limit coming from the Lyman-$\alpha$ analysis done in Ref.~\cite{McDonald:2004xn}.

Finally, we comment that the likelihood procedure given in
Ref.~\cite{McDonald:2004xn} strictly applies only to
cosmological models with a power-law matter power spectrum. This
assumption is not completely satisfied in our framework, where a
large scattering cross section (for models with $n>-4$) would
cause a cutoff in the matter power spectrum on small scales
(large $k$). In practice, as evident in Fig.~\ref{fig:ClPkdiff},
the Lyman-$\alpha$ data is restrictive enough to render the power
spectra of our models, where they are not overwhelmingly
excluded, sufficiently close in form to a simple power law in
the range of $k=\mathcal{O}(1~{\rm Mpc^{-1}})$, where reliable
data currently exists. This statement holds true for $n$ that are
not too largely positive, in which case the cutoff develops
quickly as a function of $k$; our model with $n=+2$ provides a
marginal example for this situation. For such models with large
positive velocity dependence, including $n\geq+2$, we expect our
analysis to be over-conservative, and it should be possible to
derive stronger bounds from a dedicated analysis. This situation
is analogous to that found for warm dark matter (WDM), where a
simple likelihood analysis of the type we
used~\cite{Abazajian:2005xn} finds significantly weaker
constraints than those obtained in dedicated
simulations~\cite{Seljak:2006qw}. 


\section{Baryon-DM interactions and suppression of small-scale
structure}
\label{sec:satellite_galaxies}

Baryon-DM interactions may affect small-scale
structure, and galactic substructure, in a number of ways.  The
most straightforward effect is to
suppress the growth in the early Universe of small-scale power
and thus the halo mass function at the low-mass end. Here we
estimate the effect compatible with our constraints.  We then,
in the Section that follows, discuss the possible consequences
for the evolution of galaxies at late times.

We compute the halo mass function using the extended
Press-Schechter formalism
\cite{Bond:1990iw,Bower:1991kf,Lacey:1993iv},
\be
     {dn_h\over dM}\left(M\right)={\rho_m\over
     M}\left|{d\sigma\over
     dM}\right|f\left(\delta_\chi(z),\hat\sigma\right),
\ee
where $\rho_m$ is the mean matter density in the Universe, and
$\hat\sigma$ is the variance,
\be
     \hat\sigma(M)=\int \frac{dk}{k}\Delta_m(k)\left|W(k,R)\right|^2.
\ee
Here $\Delta_m(k)$ is the matter density variance and $W(k,R)$ is a
tophat window function of radius $R$, corresponding to a halo of
mass $M=\left(4\pi R^3\rho_m/3\right)$.  For the function $f$
we use the Sheth-Tormen functional form \cite{Sheth:2001dp}, given by
\be
     f\left(\delta_\chi,\hat\sigma\right)=A{\nu\over{\hat\sigma}}\sqrt{{a\over
     2\pi}}\left[1+{1\over (a\nu^2)^q}\right]e^{-a\nu^2/2},
\ee
where $\nu=\left(\delta/\sqrt{\hat\sigma}\right)$, $a=0.75$,
$q=0.3$, $A=0.322$, and $\delta=0.686$ is the critical density
of collapse.

\begin{figure}[ht!]
\centering
\includegraphics[width=0.5\textwidth]{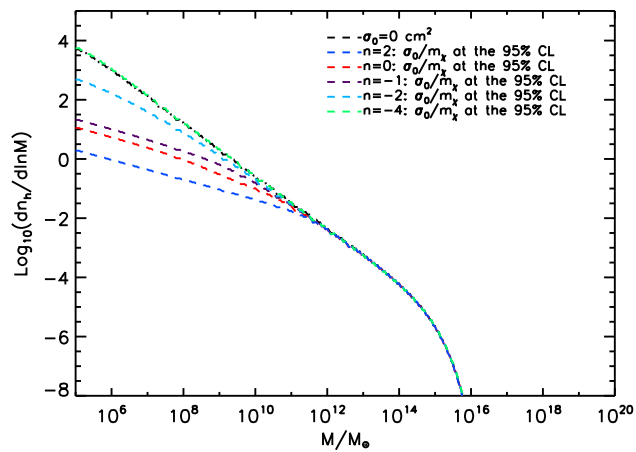}
\caption{Halo mass function as a function of mass. A model with no scattering is shown 
     in black/dashed lines, and models with different velocity-dependent cross sections are shown with a value of $\sigma_0/m_\chi$ taken at the $95\%$ CL limit
    from the analysis with CMB and Lyman-$\alpha$ data in Table~\ref{tab:1}.}
\label{fig:dnhdlnM_sigma0_vs_10GeVML_converged}
\end{figure}

Fig. \ref{fig:dnhdlnM_sigma0_vs_10GeVML_converged} shows the
halo mass function as a function of mass for  a
model with no scattering (black/dashed line) along with models
 with different velocity-dependent cross sections with a value of $(\sigma_0/m_\chi)$ taken at the $95\%$ CL limit
    from Tab.~\ref{tab:1}, using CMB + Lyman-$\alpha$ data. 
 In all cases, the cosmological parameters were fixed to the
best fit point at the given value of $(\sigma_0/m_\chi)$. 

Two main lessons can be drawn from Fig.~\ref{fig:dnhdlnM_sigma0_vs_10GeVML_converged}. First, the combined constraints from linear cosmology imply that DM-baryon scattering cannot affect the halo mass function for structures more massive than $\sim10^{12}M_\odot$. This result is model-independent. It simply reflects the scale at which the observational LSS constraint is applied in our analysis, $k\sim1$~Mpc$^{-1}$, since
\be M=\frac{4\pi}{3}\left(\frac{\pi}{k}\right)^3\rho_m\sim2\times10^{12}~M_\odot~\left(\frac{k}{1~{\rm Mpc^{-1}}}\right)^{-3}.\ee
For smaller mass halos, significant suppression of structure is in principle possible. 

Second, note that the model with $n=-4$ (scattering cross section scaling as $v^{-4}$) does not have any effect on the halo mass function. This occurs because models with $n<-3$ have the feature that they freeze-out towards high redshift, when the collision velocities (governed by thermal motion) get large, and freeze-in at lower redshift when the velocities drop. In contrast, models with $n>-3$ are initially important and then subsequently freeze-out as the Universe expands and cools. As halos of smaller mass form earlier, only modes of $n>-3$ can affect the primordial halo mass function on small scales while still satisfying CMB/LSS constraints that are only directly sensitive to $z\lesssim10^6$. 

To conclude this Section,
Fig.~\ref{fig:dnhdlnM_sigma0_vs_10GeVML_converged} teaches us
that DM-baryon scattering can affect the halo mass function on
small scales, but this effect is purely a memory effect from
early (linear) times (high redshift). This is a useful
lesson. It means that N-body simulations aiming to study the
effect of scattering, need not incorporate the scattering
explicitly. Instead, ordinary collisionless codes should be
applicable, where a modified linear matter power spectrum, as we
computed here, is used as input to encode the effect of
scattering.

Finally, we comment that the analysis of the halo mass function above is not sufficient by itself to expose highly nonlinear details within small scale objects, such as the presence or absence of central cusps etc. We briefly discuss constraints at this level of detail in Sec.~\ref{sec:discussion}.

\section{Late-time effects of baryon-DM interactions}
\label{sec:discussion}

In Sec.~\ref{sec:satellite_galaxies} we estimated the maximal effect of DM-baryon scattering on the primordial halo mass function. The effects discussed in that section encoded early time dynamics, for which our linear analysis was adequate. Here, in contrast, we comment briefly on the implications of our constraints for the late-time, non-linear evolution of galaxies.  Galaxies
are complicated objects, and the models for their detailed
structure contain considerable theoretical uncertainty (see e.g.~\cite{TrujilloGomez:2010yh}).  Still, in some generic cases of interest it is straightforward to see that our results strongly constrain the effect that DM-baryon scattering could have on the late-time evolution of galaxies. 

A clear example pertains to the halos of galaxies like our own Milky Way. The mean free path of a hydrogen atom, traversing a
typical galaxy like our Milky Way, to elastically scatter from a
DM particle is
\ba
     \label{eq:SS} 
     \lambda& \sim & 0.5 \left(\frac{\bar\sigma/m_\chi}{\rm
     cm^2/g}\right)^{-1}\left(\frac{\rho_\chi}{0.4~\rm
     GeV/cm^3}\right)^{-1}~{\rm Mpc},
\ea 
where we have used a density $\rho_\chi\simeq 0.4$~GeV/cm$^3$
characteristic of the Milky Way halo at the location of the
Solar System, and $\bar\sigma$ is evaluated for a velocity
$v\sim300$~km~s$^{-1}$~$\sim10^{-3}$, the virial velocity in the
Milky Way.  The canonical numerical value $\bar\sigma
\simeq$~1~cm$^2$~g$^{-1}$ is chosen having in mind the distance
traveled by a particle moving at $v\simeq300$~km~s$^{-1}$ over
the $10^{10}$~yr history of the Universe, $l_H\sim$~3~Mpc.
Looking at Table~\ref{tab:1}, we see that the mean-free path for
baryon-DM scattering is constrained by CMB/Lyman-$\alpha$ data
to be far larger than the distance a particle travels
through a halo in the history of the Universe: $(\lambda/l_H)\gg1$, for any $n$.  

This estimate is valid for a MW-type object at the solar cycle. To extend this estimate to larger or smaller objects like galaxy clusters and dwarf galaxies, we write
\be\label{eq:lgen}\frac{\lambda}{l_H}>30\left(\frac{\lambda_{\rm Tab.I(n)}}{\rm 100~Mpc}\right)\left(\frac{v}{10^{-3}}\right)^{-n-1}\left(\frac{\rho_\chi}{0.4~\rm
     GeV/cm^3}\right)^{-1},\ee
where for $\lambda_{\rm Tab.I(n)}$ we use the result on the last column of Tab.~\ref{tab:1}, and for $v$ and $\rho_\chi$ we use the characteristic values for the object of interest. 
We learn that $n>+2$ is needed to affect structure on galaxy cluster scales, with $\rho_\chi$ smaller by some two orders of magnitude and $v$ larger by a factor of ten or so compared to the MW halo.
In contrast, $n<-4$ is needed to affect dwarf galaxy scales, with $v\sim10^{-4}$.

\section{Conclusions}

We have provided a model-independent analysis of the constraints
imposed to baryon-DM interactions by CMB data from Planck and
Lyman-$\alpha$-forest data from the SDSS. Our work extended and improved upon earlier analyses for specific models like DM millicharge, electromagnetic dipole moments, and velocity-independent cross section (see App.~\ref{app:models}).  We found that the Lyman-$\alpha$-forest data,
included here for the first time in this context, considerably strengthen the
constraints beyond the reach of the CMB alone. We highlighted the interplay between bulk and thermal velocities, pointed out to the cross-over between them at $z\sim10^4$, and suggested an approximate way to take it into account. 

DM-baryon scattering cannot affect the halo mass function for $M>10^{12} M_\odot$. There is still room, after the new
constraints are imposed, for a potentially consequential suppression of primordial power on smaller, subgalactic scales. In the halos of galaxies like the Milky Way, outside of the innermost 1~kpc, our constraints imply that the baryon-DM interaction rate is, regardless of the model, too small
to affect the distribution of matter at late times.

Our observation that bulk DM-baryon velocities become greater
than thermal velocities at redshifts $z\simeq 10^4$ may have
interesting consequences.  It suggests that (unless the
cross-section power-law index is $n=-1$) the baryon-DM drag does
not vary linearly with the relative velocity.  If so, then the
Boltzmann equations for the evolution of baryons and dark matter
become nonlinear.  This has two consequences:  First, the
evolution of each Fourier mode cannot be described by the
standard linear equations; and second, different Fourier modes
become coupled, thus inducing non-Gaussianity.  In
this paper, we have included these effects in a mean-field
approach.  In this treatment, each Fourier mode is evolved
independently, and the coupling to other Fourier modes is taken
into account by augmenting the thermal velocity dispersion with
a dispersion due to bulk velocities.  While this approach should
be fairly accurate and provide conservative quantitative
constraints, it will be interesting in future work to
quantify these effects more precisely and to investigate the
implications of the non-Gaussianity that a baryon-DM interaction
may induce.


\acknowledgments
We thank Simeon Bird, Jo Bovy, Raphael Flauger, Anze Slosar and Matias Zaldarriaga for
useful discussions. We are particularly grateful to Uro\v s Seljak for suggesting the use of the Lyman-$\alpha$ forest data. CD was supported by the National Science
Foundation grant number AST-0807444, NSF grant number
PHY-088855425 , and the Raymond and Beverly Sackler Funds. KB was supported by DOE grant DE-SC0009988.
MK was supported by NASA NNX12AE86G, NSF 0244990, and the John
Templeton Foundation.

\appendix
\section{Thermalization}
\label{sec:thermalization}

The thermalization rate is calculated similarly to the momentum exchange rate. The change in energy of a DM particle per collision is
\ba 
     \Delta\epsilon_\chi&=&\left(\frac{\vec p_\chi+\vec
     p_b}{m_\chi+m_b}\right)\cdot\Delta\vec p_\chi=\vec
     v_{cm}\cdot\Delta\vec p_\chi,
\ea
where $\vec v_{cm}$ is the boost velocity to the center of mass frame of the collision. Focusing on the limit where peculiar velocities are smaller than the velocity dispersion, we can set the peculiar velocity to zero. The specific heating rate is then
%
\ba\label{eq:dQdt}
     \frac{dQ_\chi}{dt}&=&\frac{m_\chi\,\rho_b}{m_\chi + m_b}
     \int\frac{d\hat n_b}{4\pi}
      \int dv_bv_b^2f_b(v_b)
     \int\frac{d\hat n_\chi}{4\pi}
     \int dv_\chi
     v_\chi^2f_\chi(v_\chi)
    \int d\hat n\left(\frac{d\sigma\left(|\vec
     v_\chi-\vec v_b|\right)}{d\hat n}\right)|\vec v_\chi-\vec
     v_b|^2  \no\\
      && \times\left(\hat n-\frac{\vec v_\chi-\vec v_b}{|\vec
     v_\chi-\vec v_b|}\right)\cdot\left(\frac{m_\chi\vec
     v_\chi+m_b\vec v_b}{m_\chi+m_b}\right)\no\\
     &=&-\frac{2^{\frac{n+5}{2}}
     \Gamma\left(3+\frac{n}{2}\right)} {\sqrt{\pi}} \frac{a
     m_\chi\rho_b
     \sigma_0}{\left(m_\chi+m_b\right)^2}
     \left(\frac{T_b}{m_b} +
     \frac{T_\chi}{m_\chi}\right)^{\frac{n+1}{2}}
     \left(T_\chi-T_b\right).\nonumber \\
\ea
From this, neglecting the time derivative of the mean baryonic
molecular weight, we obtain Eq.~(\ref{eq:dotT}).

\section{Peculiar velocity beyond leading order}
\label{sec:Gn}

The exact solution for the peculiar velocity $V_\chi$ is given by
\ba
     \frac{d\vec V_\chi}{dt}&=&-\vec
     V_\chi\,\frac{c_n \rho_b\sigma_0
     \left(\frac{\dispersion}{3}\right)^{\frac{n+1}{2}}}
     {m_\chi+m_b}
     \mathcal{G}_n\left(\frac{3\vec 
     V_\chi^2}{\dispersion}\right),
\ea
in real space, with the dimensionless function:
\ba
     \label{eq:G}
     \mathcal{G}_n(w)
     &=&1+\frac{n+1}{10}w + \frac{(n+1)(n-1)}{280}w^2 
     + \frac{(n+1)(n-1)(n-3)}{15120}w^3+\cdots.
\ea 

\section{Comparison to previous work and specific dark matter-baryon interaction models}\label{app:models}

Constraints on DM-baryon scattering were derived in previous work for some specific particle physics models. Here we compare our results with existing bounds.

\subsection{Velocity-independent scattering}

Cosmological constraints on a velocity-independent scattering cross section were derived in Ref.~\cite{Chen:2002yh}. We generally agree with the derivation in~\cite{Chen:2002yh}, besides from an $\mathcal{O}(1)$ numerical difference in the expression for the momentum transfer rate $R_\chi$ and, again, from the neglect of peculiar vs. thermal velocities at low redshift. Our CMB+Lyman-$\alpha$ constraint in Tab.~\ref{tab:1} improves on the bound of Ref.~\cite{Chen:2002yh} by about two orders of magnitude, most likely due to the incorporation of Lyman-$\alpha$ data in our analysis.

\subsection{DM millicharge}
If DM carries a millicharge $q_\chi=\epsilon e$, where $e$ is the electron charge and $\epsilon\ll1$, then photon exchange with protons induces a cross section  
\be\label{eq:millis}\frac{d\sigma}{dc_\theta}=\left(\frac{m_\chi+m_H}{m_\chi m_H}\right)^2\frac{2\pi\epsilon^2\alpha^2}{v^4\left(1-c_\theta\right)^2}.\ee
We focus here on $m_\chi\gg m_H$. The scattering on helium is suppressed by the reduced mass, $\sigma_{He}=\sigma_0/4$, leading to $\mathcal{F}_{\rm He}=0.82$ when all of the helium and hydrogen are ionized. Helium recombination begins around $z\sim6\cdot10^3$ and completes by $z\sim2\cdot10^3$; it is straightforward to include helium recombination in the momentum transfer equations, but for simplicity we avoid this complication and simply set $\mathcal{F}_{\rm He}=0.76$, neglecting DM-helium scattering altogether. Note that scattering on electrons is negligible for the momentum exchange, as it amounts to $R_\chi\to R_\chi\left[1+x_e\left(1-f_{He}\right)\left(m_e/m_H\right)^{\frac{1}{2}}\right]$, where $x_e=n_e/n_H$ is the free electron fraction and $m_e=5.44\cdot10^{-4}\,m_H$ is the electron mass. 

The forward divergence in Eq.~(\ref{eq:millis}) is regulated by Debye screening due to free electrons in the plasma~\cite{McDermott:2010pa}, implying a minimum scattering angle $\theta_{\rm min}\approx2\epsilon\alpha/(3T\lambda_D)$, with $\lambda_D=\sqrt{T/(4\pi\alpha n_e)}$. This gives a momentum exchange cross section
\be\bar\sigma(v)\approx\left(\frac{m_\chi+m_H}{m_\chi m_H}\right)^2\frac{2\pi\epsilon^2\alpha^2}{v^4}\ln\left(\frac{9T^3}{4\pi\epsilon^2\alpha^3 x_en_H}\right), \ee
or in our notation (taking $m_\chi\gg m_H$, and measuring velocity in units of $c$),
\ba n&=&-4,\no\\
\sigma_0&\approx&9.6\times10^{-42}\left(\frac{\epsilon}{10^{-6}}\right)^2\left[1-0.03\,\ln\left(\frac{\epsilon}{10^{-6}}\right)\right]~{\rm cm^2}.\ea
This expression is valid for $z>1100$ or so, when $x_e\approx1$. Upon recombination, $x_e$ falls quickly below unity [$x_e(z\sim1000)\sim10^{-3}$], increasing the $\ln\epsilon$ correction to $\ln(\epsilon\sqrt{x_e})$. For the very small values of $\epsilon$ that we find here, this change in the logarithm makes no significant difference to the results. Furthermore, accounting for recombination amounts to scaling 
\be\label{eq:Rxemc} R_\chi\to {\rm min}\left(x_e,1\right) \times R_\chi,\ee
so that scattering via DM millicharge halts anyway once protons combine into neutral hydrogen.

Using the 95\%CL bound for $n=-4$ from Tab.~\ref{tab:1}, we find
\be\label{eq:mcbound}\epsilon<1.8\cdot10^{-6}\,\left(\frac{m_\chi}{\rm GeV}\right)^{\frac{1}{2}}.\ee

Cosmological constraints on DM millicharge were derived in~\cite{McDermott:2010pa,Dubovsky:2001tr,Dubovsky:2003yn,Dolgov:2013una}. Our numerical result for the 95\%CL limit is stronger by a factor of two than the bound derived in Ref.~\cite{McDermott:2010pa}. However, there are conceptual differences between our analysis and the one in~\cite{McDermott:2010pa}. Notably, we here computed the rate of momentum transfer in linear theory, while the bound derived in Ref.~\cite{McDermott:2010pa} was based on the rough argument of imposing kinetic decoupling at recombination, and ignored the bulk velocity altogether. As a result, the momentum transfer rate as defined in Ref.~\cite{McDermott:2010pa} is not the proper quantity for linear theory. In addition, as discussed in Sec.~\ref{ssec:Rs}, at redshift $z<10^4$ the typical peculiar velocity itself becomes large compared with the thermal motion. This regulates the low-velocity enhancement $(\propto v^{-4})$ of the millicharge interaction. As mentioned in Sec.~\ref{sec:CMB_constraints}, had we ignored this effect and considered only the thermal motion (as was done in Ref.~\cite{McDermott:2010pa}) we would have found a bound on $\epsilon$ that would be stronger by a factor of $\sim3$ compared to Eq.~(\ref{eq:mcbound}).

\subsection{DM electric and magnetic dipole moment}

DM magnetic and electric dipole moment (MDM and EDM) was considered in Ref.~\cite{Sigurdson:2004zp}, expressed as
\be\mathcal{L}=-\frac{i}{2}\bar\chi\sigma_{\mu\nu}\left(\mathcal{M}+\gamma_5\mathcal{D}\right)\chi F^{\mu\nu},\ee
where the MDM is $\mathcal{M}$ and the EDM is $\mathcal{D}$. 

We agree with the cosmological calculation in~\cite{Sigurdson:2004zp}. 
The momentum transfer cross section for DM-proton collisions is
\be\bar\sigma=3\alpha\mathcal{M}^2\left[1-\frac{m_H(m_H+4m_\chi)}{3(m_H+m_\chi)^2}\right],\ee
for MDM, corresponding to $n=0$ in our notation; and
\be\bar\sigma=\frac{2\alpha\mathcal{D}^2}{v^2},\ee
for EDM, corresponding to $n=-2$. 
Assuming $m_\chi\gg m_H$, ignoring helium again, and using Tab.~\ref{tab:1} we find at 95\%CL:
\ba \mathcal{M}&<&1.7\times10^{-12}~{\rm e\,cm},\\
\mathcal{D}&<&9.2\times10^{-16}\left(\frac{m_\chi}{{\rm 1~GeV}}\right)^{\frac{1}{2}}~{\rm e\,cm}.\ea

The EDM ($\mathcal{D}$) bound we find is stronger by about a factor of two than that reported in~\cite{Sigurdson:2004zp} for CMB/LSS. For MDM ($\mathcal{M}$), the original analysis of~\cite{Sigurdson:2004zp} does not apply, and our numerical result is new.


\end{document}